\documentclass[aps,superscriptaddress,floats,showpacs,floatfix]{revtex4}
\usepackage{bm}
\usepackage{graphicx}
\usepackage{subfigure}
\usepackage{amssymb}
\usepackage{amsmath}
\usepackage{color}
\usepackage[a4paper,left=1.5cm, right=1.3cm, top=3.0cm,bottom=2cm]{geometry}

\begin{document}
\title{Role of critical points of the skin friction field in formation of plumes in thermal convection}
\author{Vinodh Bandaru}
\affiliation{Institut f\"ur Thermo- und Fluiddynamik, Technische Universit\"at Ilmenau, Postfach 100565, D-98684 Ilmenau, Germany}
\author{Anastasiya Kolchinskaya}
\affiliation{Institut f\"ur Thermo- und Fluiddynamik, Technische Universit\"at Ilmenau, Postfach 100565, D-98684 Ilmenau, Germany}
\author{Kathrin Padberg-Gehle}
\affiliation{Institut f\"ur Wissenschaftliches Rechnen, Technische Universit\"at Dresden, Zellescher Weg 12--14, D-01069 Dresden, Germany}
\author{J\"org Schumacher}
\affiliation{Institut f\"ur Thermo- und Fluiddynamik, Technische Universit\"at Ilmenau, Postfach 100565, D-98684 Ilmenau, Germany}
\date{\today}

\begin{abstract}
The dynamics in the thin boundary layers of temperature and velocity is the key to a deeper understanding of turbulent transport
of heat and momentum in thermal convection. The velocity gradient at the hot and cold plates of a Rayleigh-B\'{e}nard convection cell
forms the two-dimensional skin friction field and is related to the formation of thermal plumes in the respective boundary layers. 
Our analysis is based on a direct numerical simulation of Rayleigh-B\'{e}nard convection 
in a closed cylindrical cell of aspect ratio $\Gamma=1$ and focused on the critical points of the skin friction field. We identify triplets of critical 
points, which are composed of two unstable nodes and a saddle between them, as the characteristic building block of the skin friction 
field. Isolated triplets as well as networks of triplets are detected. The majority of the ridges of line-like thermal plumes coincide with 
the unstable manifolds of the saddles. From a dynamical Lagrangian perspective,  thermal plumes are formed together with an attractive  
hyperbolic Lagrangian Coherent Structure of the skin friction field. We also discuss the differences from the skin friction field in 
turbulent channel flows from the perspective of the Poincar\'{e}-Hopf index theorem for two-dimensional vector fields.   
\end{abstract}
\maketitle

\section{Introduction}
One way to gain a deeper understanding of turbulence is to break the complex fluid motion down into a few basic 
building blocks which are connected and often appear repeatedly on different scales. Their geometry,  dynamics and interaction determines
the global statistical properties of the turbulent flow \cite{Townsend1951,Corrsin1971}. They can be identified either in the Lagrangian or Eulerian picture.
For example, the most persistent attracting and repelling Lagrangian manifolds in a turbulent flow, which are known as Lagrangian Coherent Structures 
(LCS) \cite{Haller2015}, are found to serve as barriers to the mixing of scalar 
quantities and thus affect the global scalar fluctuations \cite{Mathur2007,Mezic2010,Haller2012}. The stretching of simple Burgers-type vortices 
is used to describe the small-scale intermittency and  the related evolution of high-amplitude  enstrophy events as well as to explain 
the non-local energy transfer in the turbulent cascade \cite{Kambe1997,Lu2008,Hamlington2008,Schumacher2010}. When a passive 
scalar increment is conditioned to the distance of zero-scalar-gradient points, scaling properties similar to scalar structure functions 
follow \cite{Gibson1968,Wang2006}. In all the above examples, geometric objects are connected to dynamical processes and statistical 
properties of the corresponding turbulent flow.

In this work, we apply these ideas to the formation of thermal plumes in turbulent  Rayleigh-B\'{e}nard convection (RBC)
which is present in a fluid layer between two horizontal parallel plates heated from below and cooled from above \cite{Chilla2012}. Thermal plumes 
are fragments of the thin thermal boundary layer. They are formed in the vicinity of  the heated (cooled) plate, then rise (fall) into the bulk of the turbulent convection 
layer and provide the major contribution to the near-wall turbulent heat transfer and its local fluctuations. Their dynamics and size 
have been studied, for example, in experiments \cite{Puthenveettil2005,Zhou2007,Gunasegarane2014}  and direct numerical simulations 
\cite{Shishkina2008,Shi2012,Scheel2014,Poel2015}. Here, we will describe the formation of thermal plumes by the vertical temperature derivatives in connection to the two-dimensional velocity 
gradient vector field {\em directly at the heated (cooled) plate}. It is considered as a blueprint of the coupled temperature-velocity dynamics {\em above the heated (cooled)
plate}. The structure of the velocity gradient field at the plate is determined by its critical or zero points and their local topology. As we will see, triplets of such critical points 
play a central role in the near-wall dynamics and can be connected to the formation of line-like thermal plumes.  Their dynamics will be analyzed in the Eulerian as well
as in the Lagrangian frames of reference. We demonstrate that ridges of line-like plumes coincide with attracting hyperbolic LCS of the skin
friction field.    

The stem of thermal plumes at the hot bottom plate coincides with local minima of the absolute vertical temperature derivative $|\partial T/\partial z|$ at the plates,
where $z$ is the vertical direction. We will therefore derive and discuss a reduced transport equation 
for this vertical temperature derivative and its magnitude which is based on the fact that at the wall, the velocity gradient tensor $A_{ij}=\partial u_i/\partial x_j$ has only two 
non-vanishing components which form a two-dimensional vector field. We then relate the resulting network of critical points -- saddles and nodes  \cite{Chaos} -- 
to regions on the plate where the thermal plumes are lumped together before their take-off into the bulk. In other words, the eigenvalues which determine the phase portrait of the 
critical points appear in source-sink terms of this reduced transport equation for $\partial T/\partial z$.
 
We use the notion which has been suggested in Ref. \cite{Chong2012} and term the two-dimensional vector field, which is composed of the wall-normal
derivatives of the tangential velocity components, as the {\em skin friction field}. Critical points of the skin friction field have been studied 
recently in plane shear flows with an emphasis to understand their connection to near-wall backflows and their role in the formation of near-wall coherent 
streamwise vortices, both in simulations \cite{Chong2012,bib:card2014,bib:lena2012} and laboratory experiments \cite{Grosse2009,Bruecker2015}. It has been 
reported that critical points are partly found at the tail of an evolving large-scale structure \cite{bib:lena2012}. As we will see in the following, there are differences from the
turbulent shear flow case that has a well-defined unidirectional mean flow. On one hand, the differences arise from the different flow geometry. On the other hand, they 
also arise from the active coupling of the temperature and velocity fields which is absent in a turbulent channel flow.  

Our present work aims at applying this framework to the convection flow and plume formation. Here, we outline therefore the concepts
rather than provide an extensive parametric study with respect to Rayleigh and Prandtl number, two dimensionless parameters that characterize convective turbulence.
Such a study is certainly necessary and will be done elsewhere. This means that the concepts are tested with one turbulent convection data set at a 
moderate Rayleigh number. The analysis in the present work is done mostly at the bottom plate of the convection cell. The outline of the paper is as follows. 
We introduce in brief the Boussinesq model of convection. Afterwards, we lay out the theoretical basis
and develop a reduced equation for the evolution of the vertical temperature derivative. Our predictions are tested afterwards using high-resolution data sets at a 
moderate Rayleigh number. We close the paper with conclusions and a brief outlook. 
\begin{figure*}
\begin{center}
\includegraphics[scale=0.17]{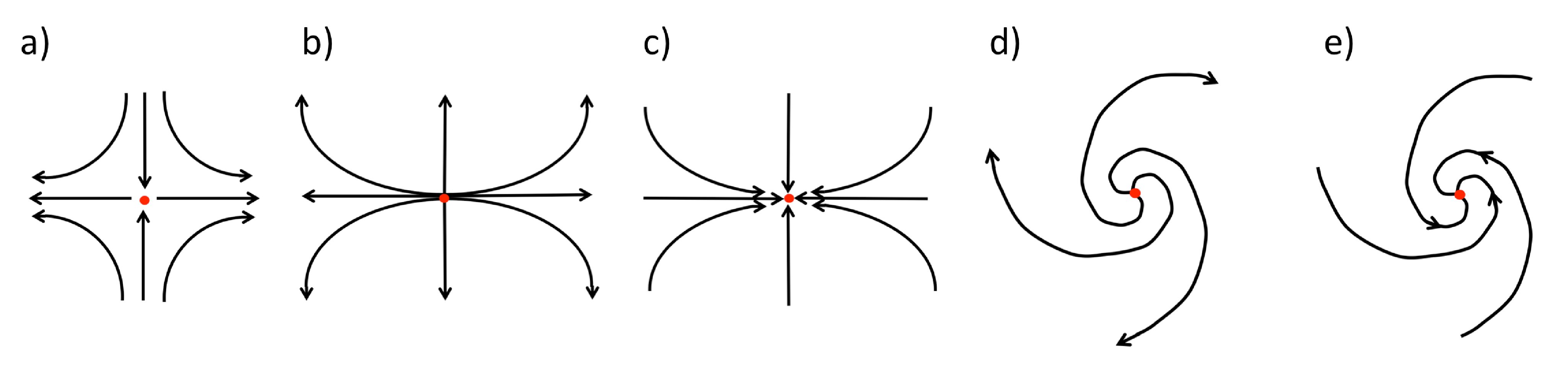}
\caption{(Color online) Critical points of a two-dimensional vector field. Phase portraits are displayed. (a) Saddle. (b) Unstable node. (c) Stable node. (d) Unstable focus.
(e) Stable focus.}  
\label{fig1}
\end{center}
\end{figure*}
  
\section{Equations of motion}
We solve the three-dimensional Boussinesq equations for turbulent RBC in a cylindrical cell of height $H$ and 
diameter $d$. The equations for the velocity field ${\bm u}({\bm x},t)$ and the temperature field $T({\bm x},t)$ are given by
\begin{align}
\label{ceq}
{\bm \nabla}\cdot {\bm u}&=0\,,\\
\label{nseq}
\frac{\partial{\bm u}}{\partial t}+({\bm u}\cdot {\bm\nabla}) {\bm u}
&=-\frac{1}{\rho_0}{\bm \nabla} p+\nu {\bm \nabla}^2 {\bm u}+ g \alpha (T-T_0) {\bm e}_z\,,\\
\frac{\partial T}{\partial t}+({\bf u}\cdot {\bm \nabla})  T
&=\kappa {\bm \nabla}^2 T\,.
\label{pseq}
\end{align}
The pressure field is denoted by $p({\bm x},t)$, the constant mass density by $\rho_0$ and the reference temperature by $T_0$ \cite{Scheel2013,Scheel2014}. The aspect ratio of the convection 
cell is $\Gamma=d/H=1$. The Prandtl number which relates kinematic viscosity $\nu$ and thermal diffusivity $\kappa$ is $Pr=\nu/\kappa=0.7$. The 
Rayleigh number is $Ra=g\alpha\Delta T H^3/(\nu\kappa)=10^7$. The variables $g$ and $\alpha$ denote the acceleration due to 
gravity and the thermal expansion coefficient, respectively. The temperature difference between the bottom and top plates is $\Delta T$. 
In a dimensionless form all length scales are expressed 
in units of $H$, all velocities in units of the free-fall velocity $U_f=\sqrt{g\alpha\Delta T H}$ and all temperatures in units of 
$\Delta T$.  We will denote the heated plate by $M$.  

We apply a spectral element method in the simulations in order to resolve the gradients of velocity and temperature accurately 
\cite{bib:nek5000}. More details on the numerical scheme and the appropriate grid resolutions can be found in Ref. \cite{Scheel2014,Scheel2013}.   
No-slip boundary conditions are applied for the velocity at all the walls. The top and bottom walls are isothermal and the side wall is thermally insulated.
The cylindrical convection cell is covered by 30720 spectral elements. On each element all turbulent fields are expanded by 11th-order 
Lagrangian interpolation polynomials with respect to each spatial direction.

\section{Theoretical basis}
\subsection{Skin friction and near-wall velocity fields}
At the bottom plate ($z=0$), the velocity gradient tensor $A_{ij}$ takes the following form
\begin{equation}
\label{tensor}
  \hat{A}\Big|_{z=0} = \left(\begin{array}{ccc}
                                0 & 0 & \partial u_x/\partial z \\
                                0 & 0 & \partial u_y/\partial z \\
                                0 & 0 & 0
                                \end{array}\right)\:.
\end{equation}
Both components form a two-dimensional wall shear stress vector field which is defined as
\begin{equation}
{\bm \tau}_w=\rho_0 \nu \frac{\partial {\bm u}_{\perp}}{\partial z}\Bigg|_{z=0}\,,
\label{wall}
\end{equation}
where the subscript ${\perp}$ denotes the two horizontal (or tangential) $x$-- and $y$--components or coordinates. 
Skin friction and wall stress fields are defined by derivatives normal to the plates. Therefore $\partial f /\partial z$ has to 
be changed to $-\partial f/\partial z$ for any field $f$ when the top plate behavior is analyzed. The skin friction field is defined  by
\begin{equation}
{\bm s}=\frac{{\bm \tau}_w}{\rho_0 \nu}\,.
\label{skin}
\end{equation}
Note that the vorticity vector at the wall, which is denoted as surface vorticity \cite{Chong2012}, is given by 
\begin{equation}
{\bm \omega}^T_{\perp}=\left(-\frac{\partial u_y}{\partial z}, \frac{\partial u_x}{\partial z}\right)\,,
\label{skin1}
\end{equation}
and is perpendicular to ${\bm s}$. In the vicinity of the top and bottom walls, we expand the velocity components as follows:
\begin{align}
{\bm u}_{\perp}({\bm x}_{\perp},z,t)&= {\bm s}({\bm x}_{\perp},t) z\,, \label{vel1a} \\
u_z({\bm x}_{\perp},z,t) &= -\frac{1}{2} ({\bm \nabla}_{\perp}\cdot{\bm s}) z^2 + C({\bm x}_{\perp})\,.
\label{vel1}
\end{align}
Again, ${\bm x}_{\perp}=(x,y)$. The integration constant $C({\bm x}_{\perp})=0$ in order to satisfy the no-slip boundary conditions 
at the wall. The velocity field which is given by Eqns. (\ref{vel1a}) and (\ref{vel1}) satisfies the incompressibility condition. It can be already seen that the essential
information of the near-wall velocity is captured by the skin friction field, in particular by its divergence. Sources and sinks of ${\bm s}$
are connected to the critical points.

\subsection{Critical points of the skin friction field}   
\subsubsection{Classification} 
The local topology or phase portrait of the critical points of the skin friction field is determined by the eigenvalues of the corresponding Jacobian which is given by
\begin{equation}
\label{tensor}
  \hat{J} = \left(\begin{array}{cc}
                                \dfrac{\partial^2 u_x}{\partial x\partial z} & \dfrac{\partial^2 u_x}{\partial y\partial z} \\
                                \dfrac{\partial^2 u_y}{\partial x\partial z} & \dfrac{\partial^2 u_y}{\partial y\partial z}                                
                                \end{array}\right)\:.
\end{equation}
We note that $\mbox{tr}(\hat{J})= ({\bm\nabla}_{\perp} \cdot {\bm s})\ne 0$. The following pairs of complex eigenvalues $\lambda_k=a_k+i b_k$ are possible when the characteristic equation is solved:
\begin{itemize}
\item Saddle: $\lambda_1=a_1<0$ and $\lambda_2=a_2>0$
\item Unstable Node: $\lambda_1=a_1>0$ and $\lambda_2=a_2>0$
\item Stable Node: $\lambda_1=a_1<0$ and $\lambda_2=a_2<0$
\item Unstable Focus: $\lambda_{1,2}=a\pm ib \;\;\;(a>0)$ 
\item Stable Focus: $\lambda_{1,2}=-a\pm ib \;\;\;(a>0)$  
\end{itemize}
The local topology of all these critical points is sketched in Fig.~\ref{fig1}.
Stable nodes or foci can be considered as sinks while unstable nodes and foci are sources. 

\subsubsection{Poincar\'{e}-Hopf theorem}
The Poincar\'{e}-Hopf theorem states that for a vector field with isolated zeros that is defined on  a compact orientable differentiable 
manifold ${\cal M}$, the Euler characteristic $\chi({\cal M})$ can be related to the sum of the indices of the $N$ isolated critical points ${\bm x}^{\ast}$ \cite{Arnold},
\begin{equation}
\sum_{k=1}^N \mbox{ind}({\bm x}^{\ast})=\chi({\cal M})\,.
\label{PHtheorem}
\end{equation}
Here, the index of a critical point ${\bm x}^{\ast}$ in a two-dimensional plane is defined as,
\begin{equation}
\mbox{ind}({\bm x}^{\ast})=\frac{1}{2\pi}\oint_{{\cal C}({\bm x}^{\ast})} d\theta\,,\;\;\;\;\;\mbox{with}\;\;\;\;\; \theta=\arctan\left(\frac{s_y}{s_x}\right)\,.
\end{equation}
Curve ${\cal C}({\bm x}^{\ast})$ surrounds the critical point. For a stable or unstable node and a focus, we find ind$({\bm x}^{\ast})=+1$, and for a saddle  ind$({\bm x}^{\ast})=-1$. 
The Euler characteristic is a number that can be assigned to the topological properties of the plane, which is the two-dimensional manifold in the following. In the wake of this theorem,
a first significant difference between turbulent channel flows and turbulent convection cells arises for simple geometric reasons.  DNS of a turbulent channel flow use typically periodic boundary conditions 
in both horizontal directions. The no-slip walls of the channel are thus two-dimensional tori and have an Euler characteristic of zero. Consequently, Eq. (\ref{PHtheorem}) implies that the number of nodes (and foci) of
the skin friction field equals the number of saddles \cite{bib:lena2012,Cardesa2014}. This is different to the present convection case. The bottom and top plates $M$
are orientable circular disks with a boundary and hence the Euler characteristic is $\chi(M)=1$. We will come back to this point when we analyze our simulation data.     

\subsection{Hyperbolic Lagrangian Coherent Structures}
As mentioned before, we denote the bottom plate of the convection cell by $M$. The time-dependent skin friction field ${\bm s}$ generates a flow map $\Phi^{t_0+\tau}_{t_0}: M\mapsto M$ which advects Lagrangian 
trajectories from time $t_0$ to $t=t_0+\tau$, i.e.,  $\hat{\Phi}_{t_0}^{t_0+\tau}({\bm x}_0): {\bm x}_0 \mapsto {\bm x}(t; {\bm x}_0, t_0)$ in correspondence with 
\begin{equation}
\dot{\bm x}={\bm s}({\bm x},t). 
\end{equation}
The Cauchy-Green strain tensor is given by
\begin{equation}
\hat{\cal C}({\bm x}_0)=\left[\nabla\hat{\Phi}_{t_0}^{t_0+\tau}({\bm x}_0)\right]^{\ast}\nabla\hat{\Phi}_{t_0}^{t_0+\tau}({\bm x}_0)\,,
\end{equation}
where  the asterisk denotes the transposed matrix. The finite-time Lyapunov exponent (FTLE) follows as 
\begin{equation}
\Lambda_{\tau}({\bm x}_0)=\frac{1}{2\tau} \log\sigma_{\tau} \;\;\;\mbox{with}\;\;\;  \sigma_{\tau}=\lambda_{max}(\hat{\cal C}({\bm x}_0))\,.
\end{equation}
Lagrangian coherent structures are the most attracting or repelling hyperbolic manifolds over a finite time interval \cite{Haller2015}. In typical situations, 
ridges in the FTLE field indicate the location of LCS \cite{Shadden2005}, so our approximation will be based on this.
In contrast to many previous applications of LCS to fluid dynamical problems, here we will examine the structures in the
two-dimensional vector field ${\bm s}$ that is not divergence free. The transport of temperature in the velocity field in the vicinity 
of the bottom plate, i.e., at $z\ll 1$, is translated to a transport of the vertical temperature derivative in the skin friction field at $z=0$. The corresponding 
transport equation is derived and simplified in the following section.

\section{Transport of wall-normal temperature derivative}
\begin{figure*}
\begin{center}
\includegraphics[scale=0.15]{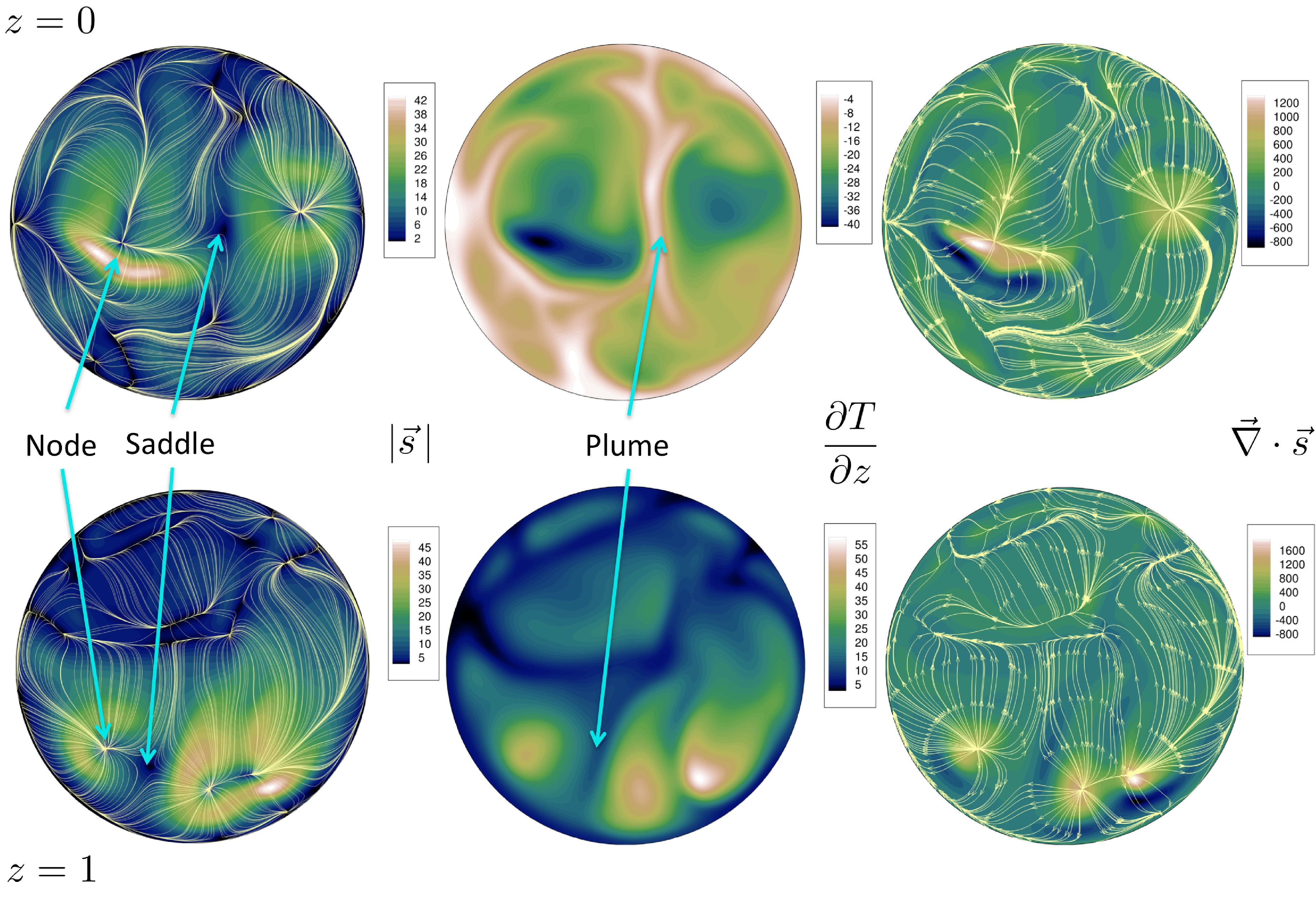}
\caption{(Color online) Simulation snapshot which displays the critical points of the skin friction field and thermal plumes. Top row: $z=0$, bottom row: $z=1$. Left: Magnitude (colored background) and streamlines
of the skin friction field ${\bm s}$. Unstable nodes and saddle point are indicated by arrows. Middle: Corresponding contour plot of the vertical temperature derivative, 
$\partial T/\partial z$. Line-like plumes are indicated again by arrows. Right: Contours of the divergence of the skin friction field together with the streamlines of ${\bm s}$.
Data are for $Ra=10^7$, $Pr=0.7$ and $\Gamma=1$.}  
\label{fig2}
\end{center}
\end{figure*}

We denote the temperature gradient by ${\bm G}=({\bm G}_{\perp},G_z)$. In the following we will derive an equation of motion for the 
vertical temperature derivative, $G_z=\partial T/\partial z$, very close to the bottom plate at $z=0$. This quantity can be directly related to the 
plumes (see also Fig.~\ref{fig2} later in the text). Line-like thermal plumes are formed 
where $|G_z|$ is very small, whereas 
regions with enhanced velocity shear will exist for large $|G_z|$, i.e., $|G_z|\gg 1$. The equation for $G_z$ follows from Eq. (\ref{pseq}).
Evolution equations for scalar derivatives have been discussed for example in Refs. \cite{Pumir1994,Brethouwer2003} in the context of passive scalar mixing.
The evolution equation for the vertical temperature derivative reads
\begin{equation}
\frac{\partial G_z}{\partial t} + ({\bm u}\cdot {\bm\nabla}) G_z=-\frac{\partial {\bm u}}{\partial z} \cdot {\bm G}+\kappa {\bm \nabla}^2 G_z\,.
\label{Gz}
\end{equation}
With the near-wall expansions (\ref{vel1}), the transport equation for the vertical temperature derivative (\ref{Gz}) close to the wall
can be rewritten as 
\begin{equation}
\frac{\partial G_z}{\partial t} + {\bm s}\cdot \left[ z {\bm \nabla}_{\perp} G_z + {\bm G}_{\perp}\right] = ({\bm\nabla}_{\perp} \cdot {\bm s}) \left[ \frac{z^2}{2}\frac{\partial}{\partial z}+ z \right ]G_z
+ \kappa  \left({\bm\nabla}^2_{\perp} + \frac{\partial^2 }{\partial z^2}\right) G_z\,.
\label{Gz1}
\end{equation}
We make several approximations near the wall and thus leave the rigorous treatment. In the spirit of an asymptotic expansion, we compare the magnitude of the different terms
in (\ref{Gz1}). Therefore, we
set $x=\tilde x H$, $y=\tilde y H$ and $z=\epsilon \tilde z H$ where the tilde stands for dimensionless quantities and $\epsilon\ll 1$. Thus 
$\partial_x$ and $\partial_y\sim 1/H$ and $\partial_z\sim 1/(\epsilon H)$. As a consequence, the horizontal contribution to the Laplacian can be 
neglected in (\ref{Gz1}). Furthermore, the two contributions to the first term on the right hand side are of the same order of magnitude and can be
summarized to one term. Close to the wall, we can additionally assume that $|G_z|\gg |{\bm G}_{\perp}|$. Together with the $\epsilon$ expansions, this simplifies (\ref{Gz1}) to
\begin{equation}
\frac{\partial G_z}{\partial t} + {\bm s}z\cdot {\bm \nabla}_{\perp} G_z \approx ({\bm\nabla}_{\perp} \cdot {\bm s})\, z G_z+\kappa \frac{\partial^2 G_z}{\partial z^2}\,.
\label{Gz2}
\end{equation}
Multiplying the equation with $G_z$ results in the following balance equation for the vertical temperature gradient magnitude
\begin{equation}
\frac{d G^2_z}{d t}  \approx 2({\bm\nabla}_{\perp} \cdot {\bm s})\, z G^2_z+2\kappa G_z\frac{\partial^2 G_z}{\partial z^2}\,.
\label{Gz3}
\end{equation}
The total time derivative on the left hand side of (\ref{Gz3}) summarizes the partial time derivative and the advection which is dominantly by the horizontal velocity components. 
If we align the coordinate axes in the vicinity of the critical points with the two eigenvectors, i.e.,  $(x=1)$ and $(y=2)$, Eq. (\ref{Gz3}) can be further simplified to
\begin{equation}
\frac{d G^2_z}{d t}\approx 2(\lambda_1+\lambda_2)\, z G^2_z+2\kappa G_z\frac{\partial^2 G_z}{\partial z^2}\,.
\label{Gz5}
\end{equation}
The first term on the right hand side of (\ref{Gz5}) is a gain term when unstable nodes or foci are locally present. In turn, the magnitude of $G_z$ decreases when stable 
nodes or foci exist. This simplified balance equation illustrates clearly the close connection between the local topology of the skin friction field which is given by the 
eigenvalues $\lambda_1$ and $\lambda_2$ and the vertical temperature derivative and thus the plumes.

As a short note, we briefly discuss the case of  free-slip boundary conditions at the heating and cooling plates which are given by
\begin{equation}
u_z(x,y,z=0)=0\,\;\;\;\mbox{and}\;\;\; \frac{\partial u_x}{\partial z}\Bigg|_{z=0}=\frac{\partial u_y}{\partial z}\Bigg|_{z=0}=0\;.
\end{equation}
In this case the skin friction field vanishes, $\bm{s}=0$, and yet the plumes are generated. In such a scenario, 
the velocity in the vicinity of the wall can be expressed to the first order as
\begin{align}
{\bm u}_{\perp}({\bm x}_{\perp},z,t)&= \bm{u}_{w}({\bm x}_{\perp},t),  \\
u_z({\bm x}_{\perp},z,t) &= -({\bm \nabla}_{\perp}\cdot{\bm{u}_{w}}) z,
\end{align}
which are the free-slip counterparts for eqns. (\ref{vel1a}) and (\ref{vel1}), respectively. Here, $\bm{u}_{w}=(u_{x},u_{y})$ represents the velocity field on the wall with sources and sinks
(see also Ref. \cite{Goldburg2001}). Using the procedure similar to that used for the no-slip boundaries, we obtain the approximate evolution equation for $G_{z}$ which is given by 
\begin{equation}
\frac{\partial G_z}{\partial t} \approx ({\bm\nabla}_{\perp} \cdot {\bm{u}_{w}})\, G_z+\kappa \frac{\partial^2 G_z}{\partial z^2}\,.
\end{equation}
It is very clear from the above equation that in the case of free-slip conditions, the divergence of the slip velocity field $\bm{u}_{w}$ acts as the source for the evolution of $G_{z}$ 
as compared to ${\bm\nabla}_{\perp} \cdot {\bm s}$ (see equation (\ref{Gz2})) in the no-slip case. In other words, with free slip boundaries, the two-dimensional slip velocity on the 
wall plays the role of skin friction field in characterizing the plume dynamics. In this paper, we only present and analyse data considering no-slip boundary conditions.
For a quantitative analysis we will now turn to the direct numerical simulation data.

\section{Simulation results}

\subsection{Skin friction field at top and bottom plates}
Fig.~\ref{fig2} illustrates the relation between the skin friction field, in particular its critical points, and the vertical temperature derivative for one time snapshot of the simulation. 
Data for the plane $z=0$ are shown in the upper row, data for $z=1$ in the lower one. The left contour plots 
display the critical points of ${\bm s}$ as the darkest regions of the contour background which stands for $|{\bm s}|$. The overlaid streamlines of ${\bm s}$ indicate the local 
topology at the critical points. We highlight a saddle and an unstable node, the latter of which are sources of the skin friction field, at both plates.
Unstable nodes are generated by cold downward flows which hit the bottom plate or hot upward flows which hit the top plate. They are surrounded by regions of high shear
as indicated by the magnitude of ${\bm s}$. The field lines of the skin friction field point almost perfectly radially away from the node, a source region of the skin friction field.     

In the mid panels we display the contours of $\partial T/\partial z$ at $z=0$ (top row) and $-\partial T/\partial z$ at $z=1$ (bottom row).  Contours of $\partial T/\partial z$ equal contours of the temperature 
$T$ slightly above the plate. This was verified, but is not displayed here.  The regions with the largest vertical
temperature derivative, i.e., with the smallest magnitude, are considered as the regions in which line-like plume ridges form and detach subsequently. They coincide 
with the regions where the magnitude of the skin friction field -- and thus the tangential stresses or the local shear -- are small. High-shear regions 
fall in between regions in which cold plumes hit the bottom plate or hot plumes are about to detach. 

The comparison of the skin friction fields at the bottom and top plates shows also that both fields are not synchronized by hot line-like plumes rising from the bottom
straight to the top and cold line-like plumes falling from the top down to the bottom of the cell. Plumes will be dispersed by the turbulence in the bulk of the cell when they rise and fall. 
The impact of colder fluid at the bottom and hotter fluid at the top, i.e., the process which generates unstable nodes of the skin friction field, is thus most probably a consequence 
of the incompressibility. The rise of plume ridges into the bulk pushes neighboring fluid back towards the bottom plate. It can be expected that this behavior depends also on the 
Prandtl number of the working fluid.        

The lower right plot of Fig.~\ref{fig2} shows the divergence of the skin friction field. Local minima of $|{\bm s}|$ can coincide with unstable nodes. This is however not a necessary 
condition as the figure shows. The data support our simple dynamical model which is condensed in Eq. (\ref{Gz5}). Critical points in the form of unstable nodes, for which 
$\lambda_1+\lambda_2>0$, are sources for $G_z^2$, i.e., the magnitude of $G_z^2$ has a local maximum in their vicinity. In these regions, cold plumes hit the 
bottom plate from above. Sinks of the skin friction field coincide with plume formation regions. When $\lambda_1+\lambda_2<0$, the first term in (\ref{Gz5}) turns into a loss term for $G_z^2$.  

\subsection{Derivative statistics}
\begin{figure}
\begin{center}
\includegraphics[scale=0.45]{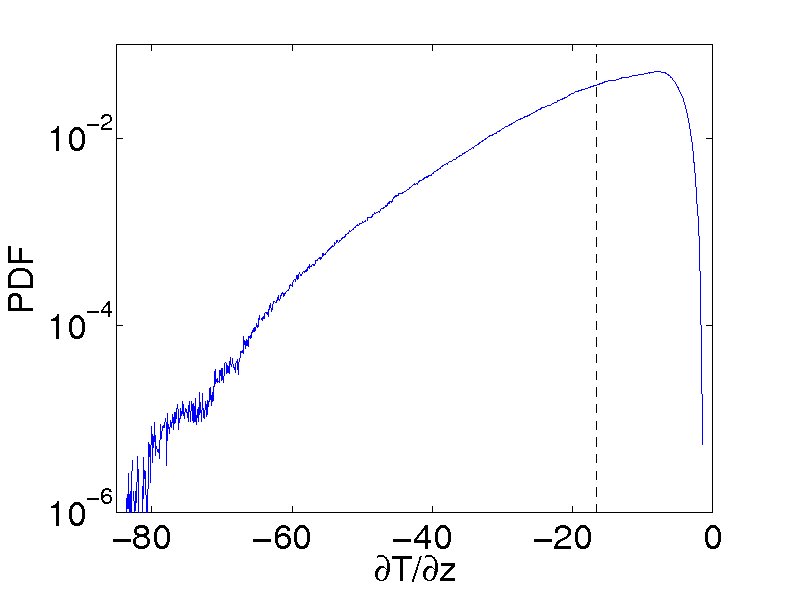}
\caption{(Color online) Probability density function of the vertical temperature derivative $\partial T/\partial z$ which is taken at the 
bottom plate, $z=0$. Data are for $Ra=10^7$, $Pr=0.7$ and $\Gamma=1$. The dashed vertical line indicates the mean vertical temperature at the 
bottom plate which equals the Nusselt number of the convection flow.}  
\label{fig3}
\end{center}
\end{figure}

\begin{figure}[!h]
        \subfigure[]{
                \includegraphics[width=0.47\textwidth]{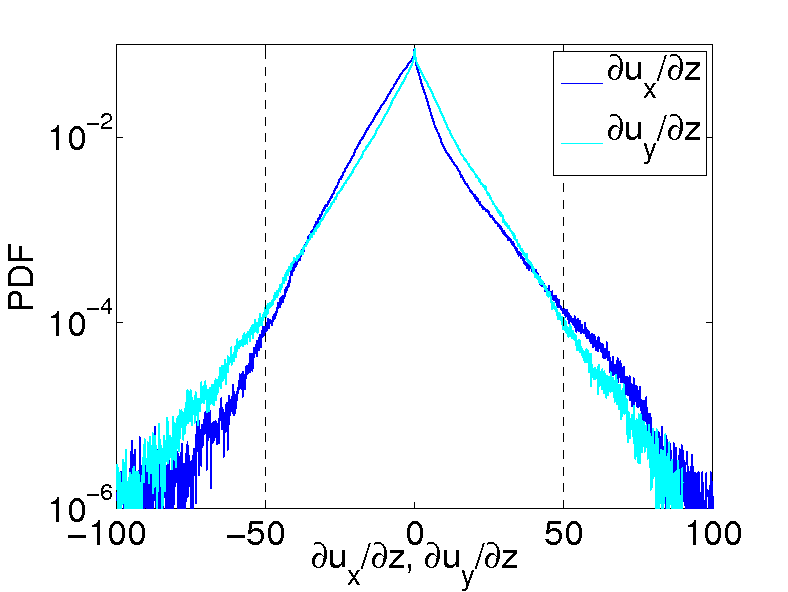}
                \label{fig4a}
        }
        \subfigure[]{
                \includegraphics[width=0.47\textwidth]{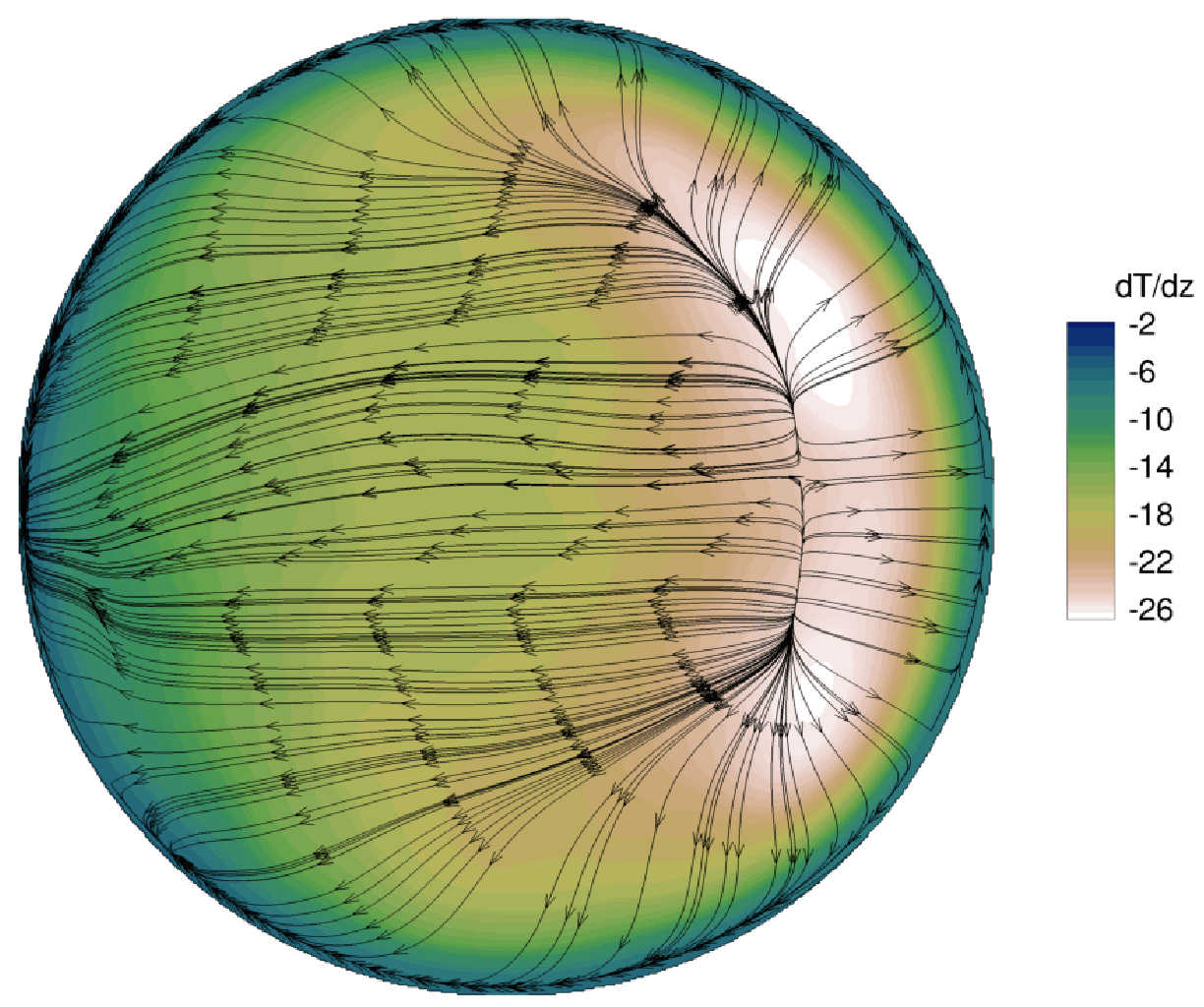}
                \label{fig4b}
        }
\caption{(Color online) a) Probability density function of the two components of the skin friction field, $\partial u_x/\partial z$ and $\partial u_y/\partial z$ at $z=0$.
The data set for this analysis is the same as in Fig.~\ref{fig3}. Mean values are $\bar{s}_{x}=-3.12$ and $\bar{s}_{y}=-0.49$ respectively. b) Lines of mean skin friction field
averaged over 202 time snapshots which are separated by roughly one free-fall time unit $T_f=H/U_f$ of each other. The background contour stands for $\partial T/\partial z$.}
\label{fig4}
\end{figure}

The vertical derivatives of the velocity and temperature are discussed in Fig.~\ref{fig3}.   
In this figure, we display the probability density function (PDF) of the vertical temperature derivative at $z=0$. It has negative amplitudes only and develops an extended 
tail. Data have been obtained from 230 statistically independent samples of the turbulent convection flow which have been gathered along a long-time integration.
They are separated by at least one free-fall time $T_f=H/U_f$. Figure~\ref{fig4a} display the statistics of the two components of the skin friction field 
obtained from the same dataset. The whole bottom plate is used for the analysis in the following. We verified that the exclusion of side wall boundary data does not change 
the results. The distributions display the typical exponential tails which reflect the intermittent 
statistics of velocity derivatives. 

It is interesting to note that the mean values of the distribution differ ($\bar{s}_{x}=-3.12$ and $\bar{s}_{y}=-0.49$)
indicating a large scale circulation (LSC) which breaks azimuthal symmetry. This is readily confirmed from the streamlines of the time-averaged skin friction field shown for $z=0$ 
in Fig.~\ref{fig4b}, that are mostly aligned in the negative $x$-direction. Accordingly, we see an impact region on the right and a ``take off`` region on the left.
In addition, the left tail of the PDF of $s_{x}$ is seen to be systematically fatter than that of $s_{y}$ in the range when $|s_{x}|<40$ and the opposite ocuring
for the right tails. It is known from other DNS studies \cite{Shi2012,Emran2015} that LSC exhibits a very slow angular drift (at time scales larger than 1000 free-fall units $T_f$), which explains
the biased directionality of the LSC in our case due to the fact that the time window for averaging was smaller.
In addition, the LSC depends on $Ra$ and becomes less coherent at large $Ra$. The Rayleigh number of $10^7$ in our case being relatively small, we thus see a clear indication of the LSC.

\begin{figure}
\begin{center}
\includegraphics[scale=0.45]{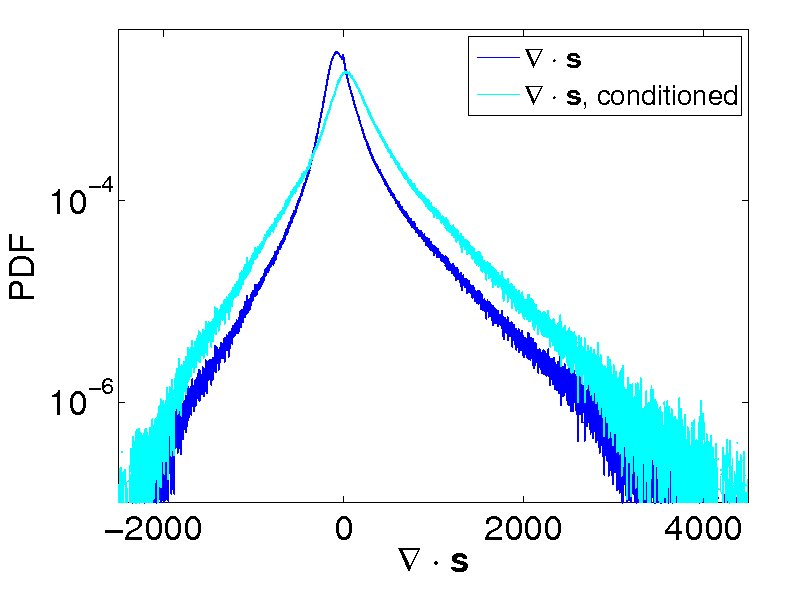}
\caption{(Color online) Probability density function of the divergence of the skin friction field. We compare the unconditioned divergence with the 
divergence for which $\partial T/\partial z< \langle \partial T/\partial z\rangle_{A,t}$. The data set for this analysis is the same as that in Fig.~\ref{fig3}.}  
\label{fig5}
\end{center}
\end{figure}
As a next step, we connect the statistics of the vertical temperature derivative with that of the skin friction field.  
Figure~\ref{fig5} shows the statistics of ${\bm \nabla}\cdot {\bm s}$. While one graph displays the unconditioned divergence field, the other graph is
conditioned to the highest magnitude temperature derivatives, i.e. to events with $\partial T/\partial z<\langle \partial T/\partial z\rangle_{A,t}$ where 
$\langle\cdot\rangle_{A,t}$ is a plane-time average taken at $z=0$. The whole positive tail of the conditioned PDF exceeds that of the 
unconditioned PDF. This finding statistically underlines the observation, which was discussed earlier: positive divergence of the skin friction field is 
connected with unstable nodes which hit the bottom plate and generate high magnitude temperature derivatives.

\subsection{Eigenvalue analysis of critical points}
The critical points of the two-dimensional skin friction field are usually not located on the grid points where all the discretized field variables are approximated. Due to this reason, 
a simple search for them `on' the grid points is insufficient. To determine the critical points, the following procedure is adopted which is similar to the observation that was discussed in Cardesa et al. \cite{Cardesa2014}. 
At each grid point $k$, considering the local gradients of ${\bm s}$,
the position $\bm{dr}^{k}=\left[ dx^{k},dy^{k}\right]$ (relative to the grid point being considered) at which ${\bm s}=0$ is computed by solving
\begin{equation}
\hat{J}^k \bm{dr}^k= -{\bm s}^{k}
\end{equation}
where the superscript $k$ on $\hat{J}$ and ${\bm s}$ indicates values computed at the grid point $k$. Any point so obtained, which is farther than $\delta$ from the grid
point $k$ is discarded (i.e. if $|\bm{dr}^{k}|>\delta$, $\delta$ being the grid size) and we denote the remaining points (which are `potential' critical points) as the set $\bm{P}$. 

In the next step, all the points in $\bm{P}$ are traversed to identify `clusters' of points, where a cluster refers to a group of points that are within a distance $\sqrt{2}\delta$
from each other. In this process, points that do not have at least a single neighbor are discarded as outliers. Typically each cluster contains $2$ to $4$ points which are located within a
grid cell and were correspondingly obtained from the corner grid points of the cell. Finally, from each of these clusters one point is chosen randomly to qualify as a critical point and the others discarded.
In this way, it is ensured that all the critical points are identified accurately. The type of the critical point is determined from the characteristics of the eigenvalues of $\hat{J}$ as 
described in subsection III. C. For this analysis, the fields on the unstructured spectral element grid have been interpolated to a uniform Cartesian mesh, which simplifies the search 
of critical points significantly.     
\begin{figure}
\begin{center}
\includegraphics[scale=0.2]{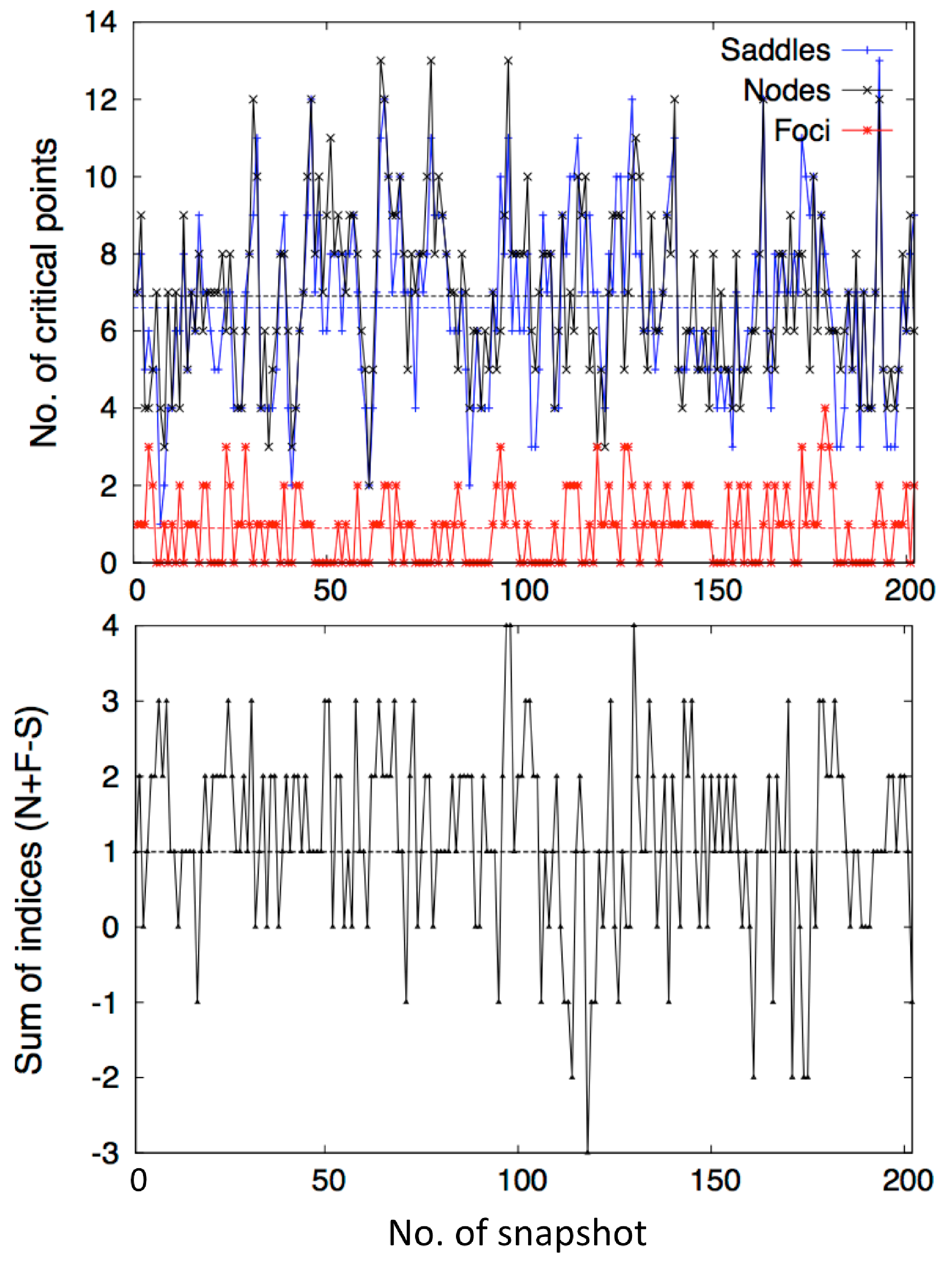}
\caption{(Color online) Top: Number of saddles, nodes and foci for the individual snapshots of the turbulent convection flow. The dashed lines with same color as the data
correspond to the arithmetic means. Bottom: Sum of the indices of saddles and nodes. The dashed line at 1 indicates the Euler characteristic for the hot bottom or cold top plates $M$. 
Same data are taken as in top figure.}  
\label{fig6}
\end{center}
\end{figure}

In Fig.~\ref{fig6} (top) we display the results of the eigenvalue analysis using 202 statistically independent time snapshots of the DNS. 
The number of foci is small compared to the number of saddles and nodes. It can be observed that the number 
of critical points is strongly fluctuating. Furthermore, we detect that the number of nodes and saddles is not the same (see bottom panel of Fig.~\ref{fig6}). This is a significant difference from the 
numerical studies in the channel flow. The reasons are as follows. First, the Euler characteristic is not zero for our circular disk with boundary as we have discussed in Sec. III B. Secondly, due to the no-slip boundary 
conditions, the skin-friction field vanishes on the boundary, i.e., ${\bm s}=0$, so that the Poincar\'e-Hopf theorem for manifolds with boundaries is not applicable. Thirdly, for the same reason 
even the critical points in the interior of the disk are often found very close to sidewalls. We note that an extension of the Poincar\'e-Hopf theorem \cite{Ma2001} holds also for the case of 
isolated critical points on the boundary of $M$. Thus the search algorithm for the critical points excludes a very tiny region at the boundary of the circular bottom and top plate, in order to 
restrict to saddles and nodes away from the sidewall. Only these critical points are in the focus of this analysis and remain an important structural element of the skin friction field of thermal convection compared 
to channel flows. 
\begin{figure}
\begin{center}
\includegraphics[width=0.6\textwidth,trim=30mm 20mm 20mm 30mm,clip]{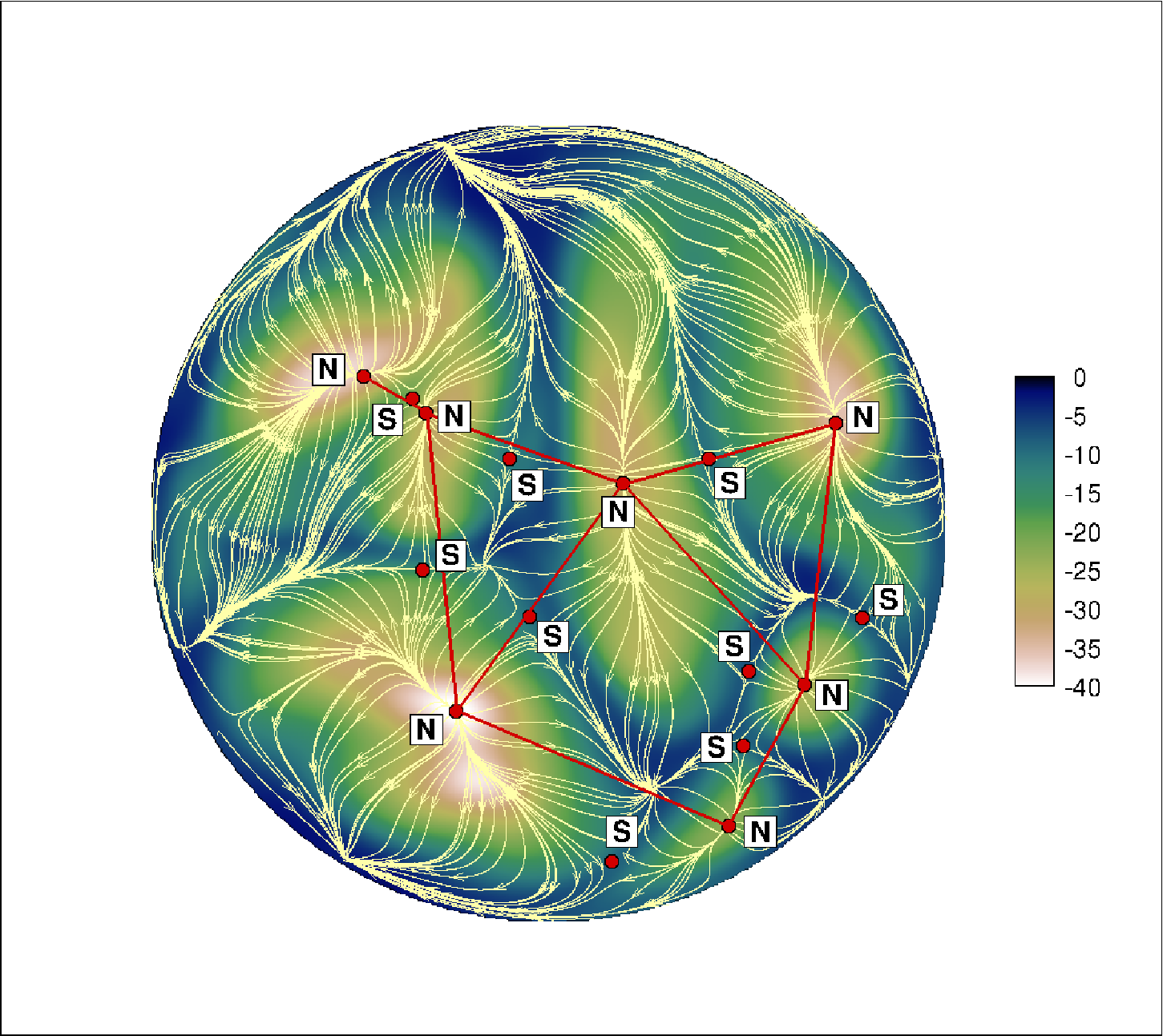}
\caption{(Color online) Network of critical points formed by N-S-N triplets of saddles (S) and nodes (N) at $z=0$. A snapshots of the skin friction field is shown. 
The solid lines between the nodes are thought as a guide to the eye. Saddles are always found in between two nodes close to this straight connection. The contours represent the
temperature derivative $\partial T/\partial z$.} 
\label{fig7}
\end{center}
\end{figure}
\begin{figure}
\begin{center}
\includegraphics[width=0.6\textwidth,trim=30mm 20mm 20mm 30mm,clip]{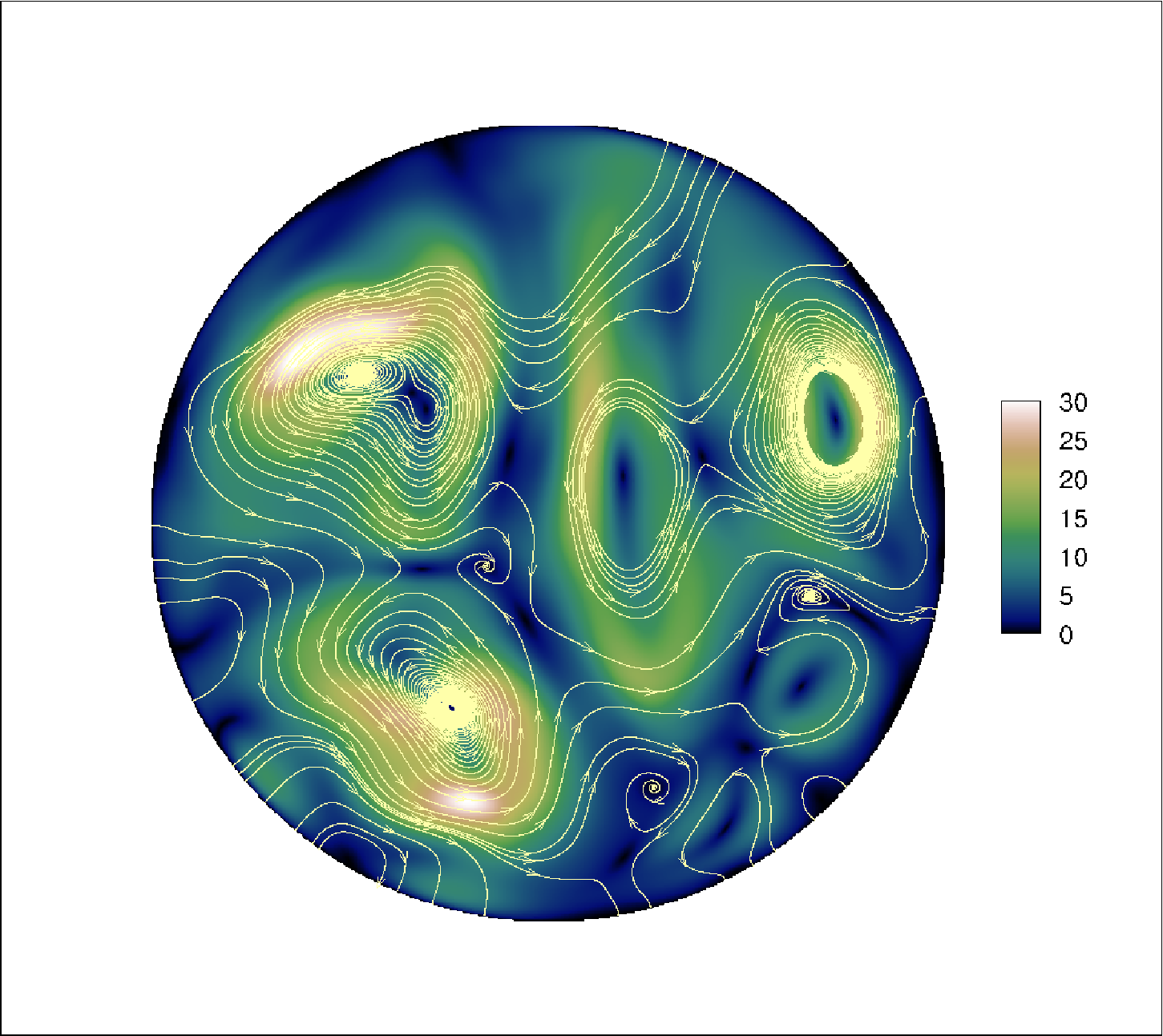}
\caption{(Color online) Field lines of the surface vorticity field plotted together with the magnitude of the skin friction field, $|{\bm s}|$, as 
background contour plot. The data are the same as in Fig.~\ref{fig7}.}  
\label{fig8}
\end{center}
\end{figure}

Figure~\ref{fig7} underlines that the plume ridges can be associated in most instances with node-saddle-node 
triplets. These triplets can be found isolated as in Fig.~\ref{fig2}, or they can form a whole triplet network as shown in this figure. Fluid is expelled radially outward from the unstable nodes 
which are distributed over the plate. Centered between two such nodes is a saddle point of ${\bm s}$ which is formed in a stagnation point flow configuration. A similar behavior is observed
in other snapshots which we have analyzed in detail. It is needless to say that exactly the same dynamics is present at the top plate for the formation of cold plumes that fall into the bulk of the cell.

How can the regions around unstable nodes be separated from each other? Figure~\ref{fig8} displays field lines of the surface vorticity,  ${\bm \omega}_{\perp}$, for which ${\bm \omega}_{\perp}\cdot{\bm s}=0$ 
holds. By comparing this figure and Fig.~\ref{fig7}, one can see that the unstable nodes of the skin friction field are surrounded by closed field lines of surface vorticity. The outermost closed 
field lines of the surface vorticity coincide with the local regions of enhanced shear which form between the regions where the cold (hot) plumes hit the bottom (top) plate and where new plume ridges are 
formed. This outermost closed field line is in some cases detected as a limiting streamline to which closed field lines of the surface vorticity converge from the in- and  outside. This is observed for the upper left node 
in Fig.~\ref{fig8} and is connected with field lines of the skin friction field that remain not perfectly radial further away from the critical point.
    
\subsection{Lagrangian analysis of line-like plume formation}
The Lagrangian analysis takes into account the \emph{time-dependent} dynamics and is based on a set of 300 snapshots which are separated by $dt=0.014 T_f$ from each other. The data have been 
obtained at the same $Ra$, $Pr$ and $\Gamma$. The integration time step is $dt/20$ and thus a total time of $t=4.2 T_f$ is spanned by this data set. The fine time resolution is necessary to conduct the 
Lagrangian analysis in order to determine the LCS. We integrate Lagrangian trajectories initialized on a fine grid by a fourth-order Runge-Kutta method and interpolate linearly with respect to time 
between the snapshots and by means of cubic splines with respect to the space. Approximations of the FTLEs are obtained by computing the maximum relative dispersion between neighboring 
trajectories, see e.g.~\cite{Padberg2007}. The LCS are then extracted from the local maxima of the resulting scalar field.
Figure 9 displays the result for three time instants $t_i$. Here attracting LCS are obtained by a backward-in-time integration starting at $t_{0,i}=t_i-10dt$.  They are 
highlighted as the black solid lines in the top row contour plots of $\partial T/\partial z$. For comparison, we replot in the bottom row of the same figure the streamlines of the skin friction field together
with the node-saddle-node triplets. The result confirms that the attracting LCS which are largely determined by the unstable manifolds of the saddles in the time-frozen fields 
are found at the line-like plume ridges.

\begin{figure*}
\begin{center}
\includegraphics[scale=0.17]{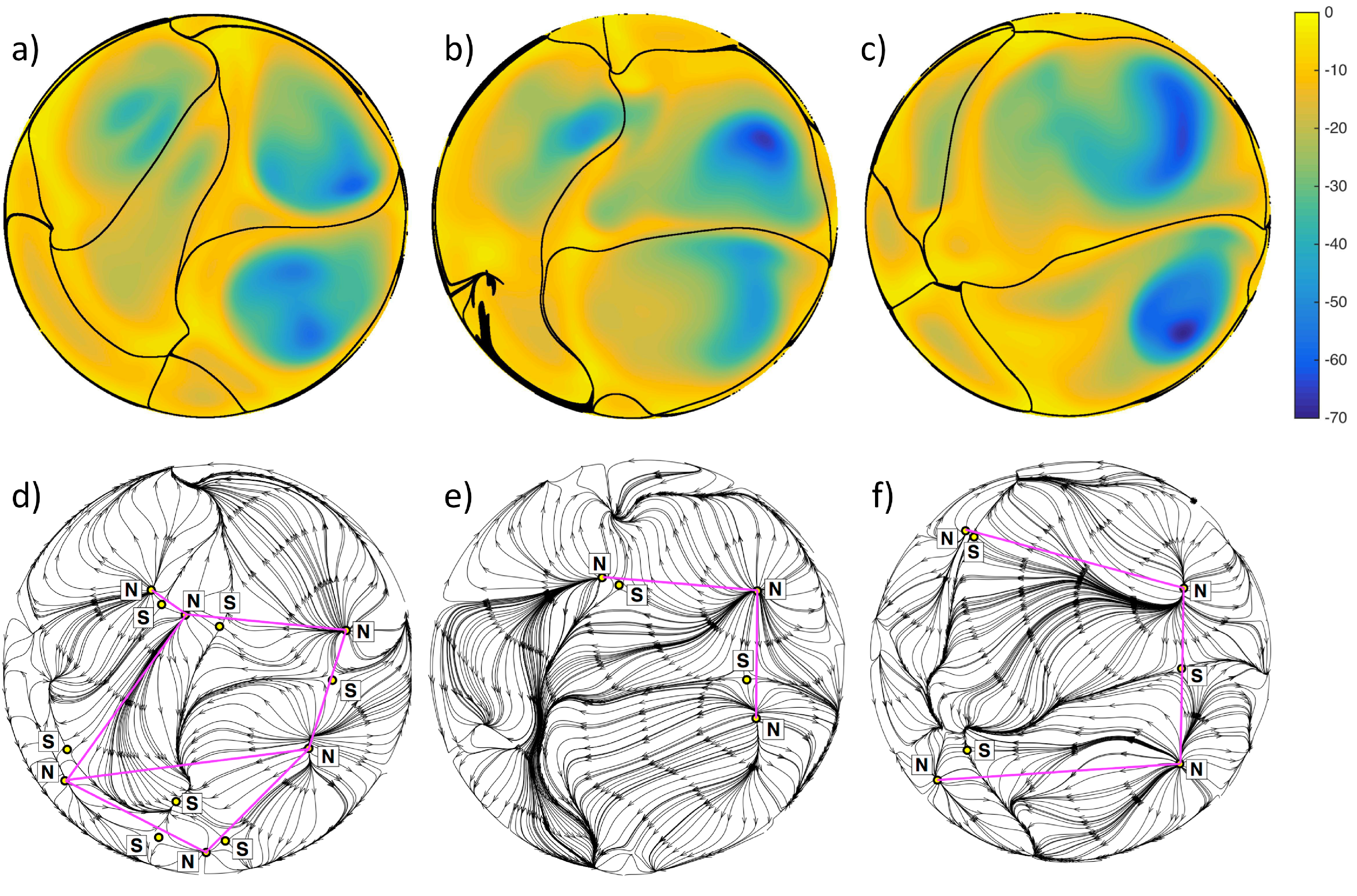}
\caption{(Color online) Comparison of Lagrangian analysis (a-c) and corresponding Eulerian analysis (d-f).  Three snapshots of the skin friction field dynamics are shown. The top row shows the 
attracting Lagrangian Coherent Structures as solid lines. The background shows the vertical temperature derivative. In the bottom row, field lines of the skin friction field and the node-saddle-node triplets are 
displayed.}  
\label{fig9}
\end{center}
\end{figure*}

\section{Summary and discussion}
We discussed the connection between the moving skeleton of critical points of the skin friction field and the vertical temperature derivative {\em at the heated plate} as a blueprint of the near-wall 
dynamics of line-like plume formation {\em right above the heated plate}. Node-saddle-node triplets are identified as the typical configurations which form line-like plumes. Nodes are found where cold fluid 
hits the bottom plate. This is in contrast to former results in plane shear flows in which saddles and nodes form almost exclusively pairs rather than triplets. Beside the different geometry of the plates, 
there are several further differences between a boundary layer in a convection cell and a plane channel. For the RBC case, the unidirectional mean flow is absent. As it was discussed in Refs. 
\cite{Ahlers2009,Chilla2012}, the mean flow in a convection cell is fully three-dimensional and obeys its own complex slow dynamics (see also \cite{Emran2015}). Fluid turbulence is generated by buoyancy 
forces in the present case. Line-like thermal plumes, which form a skeleton that moves across the plate, are a typical feature that does not exist in a shear flow. The plume ridges enclose regions in which 
cold fluid is pushed back towards the bottom plate and generates unstable nodes of the skin friction field. Ring-like regions of high shear form around these unstable nodes. They can be identified by closed field 
lines of the surface vorticity. 

We have also discussed that our observation does not violate the Poincar\'{e}--Hopf theorem since $\chi(M)=1$ for the present case.  Finally, the application of Lagrangian 
Coherent Structure analysis showed that our ideas hold in a dynamical framework. The unstable manifolds of the saddles points visually coincide with attracting hyperbolic Lagrangian Coherent Structures. 
Attracting hyperbolic LCS are also found where the plume ridges are formed which subsequently rise into the bulk.

With a view to the future work several points are interesting. First, the local flow reversals which should be found in the vicinity of foci are rare as Fig.~\ref{fig6} shows. This is similar to the channel flow 
analysis. Very recently, they have also been found experimentally \cite{Bruecker2015} in a shear flow and are thought to play a central role for the formation of large-scale structures. Their role in RBC is yet 
open since the large-scale flow is not uni-directional and depends on time. The percentage and dynamical relevance of foci could increase as the Rayleigh number grows.

Second, we studied here one data set in order to demonstrate the ideas and concepts. It is certainly interesting and necessary 
to verify how robust the picture is for higher Rayleigh numbers. Furthermore, what happens when the Prandtl numbers are much larger or smaller than one and both boundary layers get decoupled?  
It can be expected that the dynamics is very different for the high-Prandtl case in comparison to the low-Prandtl case (at same $Ra$) simply due to the fact that the coherent large-scale motion
is significantly weaker in the former case. Some of these analyses have been started very recently and will be reported elsewhere.  

Third, we saw that the skin friction fields at the top and bottom plates are not directly synchronized.  The reason is that thermal plumes cannot rise straight to the top plate for the present Prandtl numbers.
However, a large-scale circulation exists in a closed cell. For the present aspect ratio, a single large roll is expected \cite{Bailon2010} which couples the dynamics at the top and bottom plates. The particular 
way of coupling requires further detailed analysis.  

\begin{figure*}[!h]
\begin{center}
\includegraphics[scale=0.12]{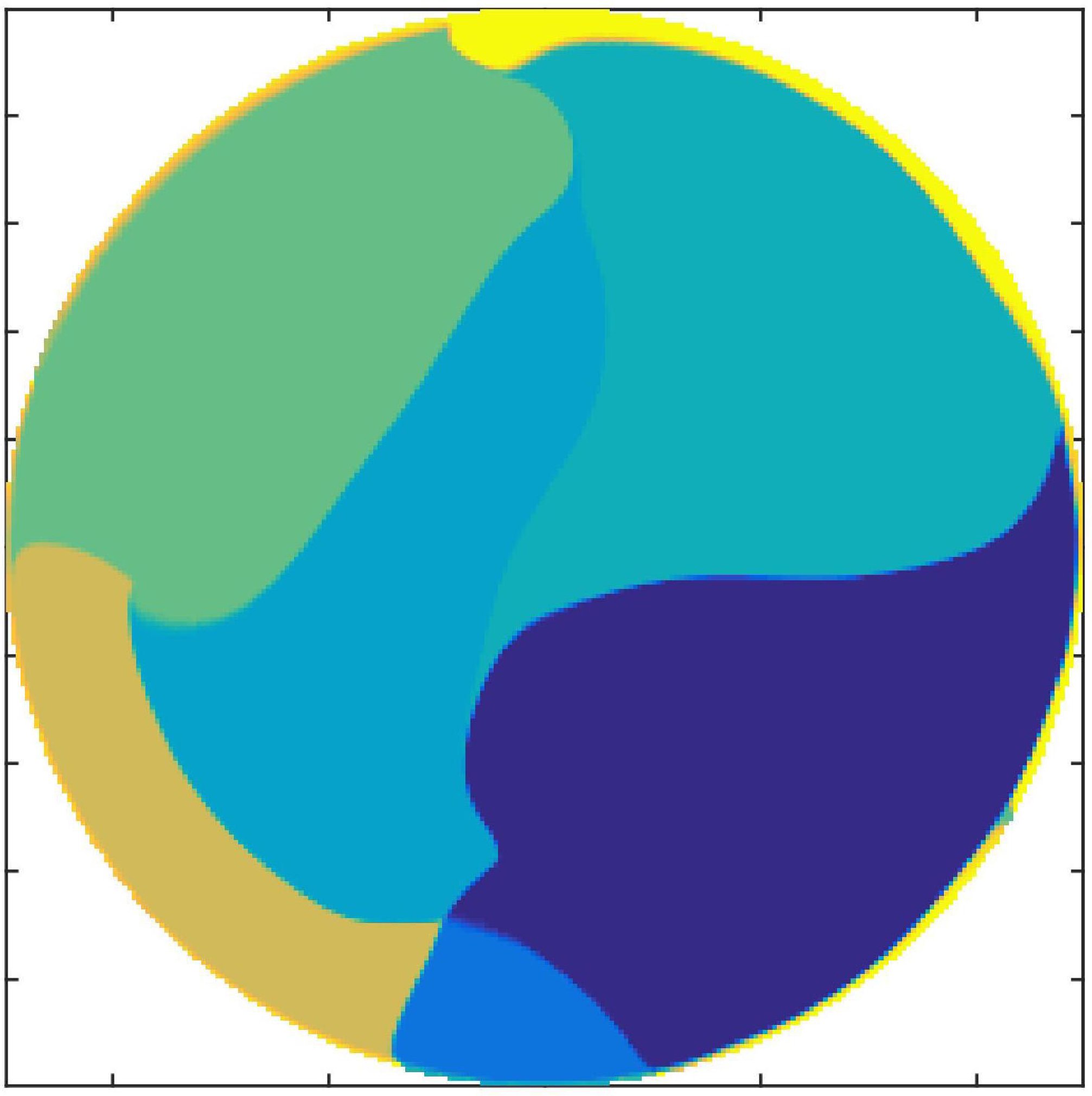}
\caption{(Color online) Direct numerical approximation of the regions around unstable nodes of the skin friction field. Dominant eigenvectors of a stochastic matrix sampled from Lagrangian trajectories highlight 
the dynamically distinct regions, which compare well with the corresponding results for the Lagrangian coherent structures in Fig.~\ref{fig9}(a).}   
\label{fig10}
\end{center}
\end{figure*}     

Finally, as especially the regions around unstable nodes are key to the plume formation, a direct and robust 
approximation is needed for further investigation. Set-oriented numerical methods using stochastic matrices are tailored for finding dynamically distinct regions in complex flows \cite{Dellnitz1999,Froyland2014}. 
A preliminary study along this line is shown in Fig.~\ref{fig10}. The resulting partition of the bottom plate compares well with the corresponding LCS which is shown for the same data set in Fig. \ref{fig9}(a). 
The application of such transfer-operator based approaches to the computational analysis of plume formation in the full three-dimensional Rayleigh-B\'{e}nard system is a subject for future research.
In order to quantify for example the transport of heat and momentum which is associated with the sets that are shown in Fig.~\ref{fig10}, extensions of the existing mathematical framework will be necessary.

\acknowledgements
The work of VB and AK (partly) is supported by the Research Training Group GK 1567 which is funded by the Deutsche Forschungsgemeinschaft.
AK is also supported by Grant SCHU1410/18 of the Deutsche Forschungsgemeinschaft.
Computing resources have been provided by the John von Neumann Institute for Computing at the J\"ulich Supercomputing 
Centre under Grant HIL07. We are grateful for this support. JS thanks Bruno Eckhardt for discussions on this subject and Philipp
Schlatter for his initial help with the data interpolation.

\end{document}